\documentclass[12pt]{article}
\usepackage{epsfig}
\usepackage{latexsym}
\begin{document}
\title{Exhaustive Ghost Solutions to Einstein-Weyl Equations for Two Dimensional Spacetimes}
\author{\begin{tabular}{c}\bigskip Muxin Han$^{1}$\footnote{Email: hamsyncolor@hotmail.com}, Yapeng Hu$^{1}$\footnote{Email: huzhengzhong2050@163.com}, Hongbao Zhang$^{1,2}$\footnote{Email: hbzhang@pkuaa.edu.cn}\\  \smallskip$^1$Department of Physics, Beijing Normal
University, Beijing, 100875, PRC
\\ \smallskip$^2$Institute of Theoretical Physics, Academia Sinica,
Beijing, 100080, PRC \end{tabular}} \maketitle
\begin{abstract}
Exhaustive ghost solutions to Einstein-Weyl equations for two
dimensional spacetimes are obtained, where the ghost neutrinos
propagate in the background spacetime, but do not influence the
background spacetime due to the vanishing stress-energy-momentum
tensor for the ghost neutrinos. Especially, those non-trivial
ghost solutions provide a counterexample to the traditional claim
that the Einstein-Hilbert action has no meaningful two dimensional
analogue.
\end{abstract}
\section{Introduction}
It is known that ghost solutions to Einstein-Weyl equations have
been obtained for various cases\cite{Griffiths,CM,DR,TDC}, where
the stress-energy-momentum tensor for the ghost neutrino field
vanishes although the field gives rise to a non-vanishing current.
Obviously, the most interesting thing about the ghost solutions is
the way the neutrino field propagates in the background spacetime
without changing it, which implies we could not detect the ghost
neutrinos by their gravitational effects. In addition, this kind
of ghost solutions may be useful to quantum field theory in curved
spacetimes, where the neutrino field could be quantized while
leaving the background spacetime fixed\cite{DR1}.

The present work is devoted to exhaustive ghost solutions to
Einstein-Weyl equations for two dimensional spacetimes, which has
never been investigated so far, as we know. It is worth noting
that our result presents a counterexample for the traditional
belief that Einstein-Hilbert action makes no sense in two
dimensional case\cite{Strobl}.

This paper is organized as follows. Section 2 investigates the
conformal invariance of  Einstein-Weyl equations for two
dimensional spacetimes. In Section 3, exhaustive ghost neutrino
solutions in two dimensional flat spacetimes are obtained. Section
4 contains some concluding remarks and discussions. The derivation
of Einstein-Weyl equations by variational principle is relegated
to Appendix.
\section{Conformal Invariance of Einstein-Weyl Equations for Two
Dimensional Spacetimes}
Start with two metrics conformally related
by
\begin{equation}
\widetilde{g}_{ab}=\Omega^2g_{ab},
\end{equation}
then for convenience but without loss of generalization, we let
\begin{equation}
{\widetilde{\sigma}^b}_{CC'}=\Omega^{-1}{\sigma^b}_{CC'}.
\end{equation}
Later, through a routine computation and by virtue of
(\ref{lemma1a}), the corresponding ${{\Theta_a}^B}_C$ in Appendix
reads
\begin{equation}
{{\Theta_a}^B}_C=\frac{1}{2}{\epsilon_C}^B\bigtriangledown_a\Omega-\sigma_{aCC'}\sigma^{dBC'}\bigtriangledown_d\Omega.
\end{equation}
Thus if we let
\begin{equation}
\widetilde{\psi}^B=\Omega^m\psi^B,
\end{equation}
then
\begin{equation}
\widetilde{\bigtriangledown}_a\widetilde{\psi}^B=\Omega^m\bigtriangledown_a\psi^B+\Omega^{m-1}[(m+\frac{1}{2})\psi^B\bigtriangledown_a\Omega-\psi^C\sigma_{aCC'}\sigma^{dBC'}\bigtriangledown_d\Omega].
\end{equation}
According to the above identity and employing (\ref{lemma1b}), it
is easy to know, for two dimensional case, the Weyl neutrino field
equation is conformally invariant with conformal weight
$m=-\frac{1}{2}$; furthermore, the stress-energy-momentum tensors
for Weyl neutrino field are conformally invariant in the sense
that
\begin{equation}
\widetilde{T}_{ab}=T_{ab}.
\end{equation}
In addition, the Einstein tensor is also conformally invariant for
two dimensional spacetimes. Thus for two dimensional spacetimes
the Einstein equation is conformally invariant. Furthermore,
according to the well-known fact, that is, any two dimensional
spacetime metric is locally conformal to a flat one, we know the
Einstein tensor vanishes identically. Hereby there is no existence
of solutions to Einstein-Weyl equations for two dimensional
spacetimes except the ghost solutions under consideration, which
are conformally invariant.

Therefore, our task is reduced to searching of ghost neutrino
solutions in two dimensional flat spacetimes.

\section{Exhaustive Ghost Neutrino Solutions in Two Dimensional Flat Spacetimes}
Given a two dimensional flat spacetime
\begin{equation}
ds^2=dt^2-dx^2.
\end{equation}
Perform a coordinate transformation
\begin{eqnarray}
v&=&t+r,\nonumber\\
u&=&t-r;
\end{eqnarray}
then
\begin{equation}
ds^2=dudv.
\end{equation}
Construct the null tetrad-frame in N-P formalism as
\begin{eqnarray}
k&=&\sqrt{2}\frac{\partial}{\partial v},\nonumber\\
l&=&\sqrt{2}\frac{\partial}{\partial u};
\end{eqnarray}
then the corresponding spin coefficients all vanish. Therefore the
Weyl neutrino field equation reads
\begin{equation}
D\psi^{1}=0,\label{weyl1}
\end{equation}
\begin{equation}
\Delta\psi^{2}=0;\label{weyl2}
\end{equation}
which implies
\begin{equation}
\psi^{1}=\psi^{1}(u),
\end{equation}
\begin{equation}
\psi^{2}=\psi^{2}(v).
\end{equation}
Thus the vanishing stress-energy-momentum tensor for ghost
neutrinos can be expressed as
\begin{equation}
T_{00}=-\frac{\sqrt{2}\alpha_N}{32\pi}(\overline{\psi}^2\frac{\partial\psi^2}{\partial
v}-\psi^2\frac{\partial \overline{\psi}^2}{\partial
v})=0,\label{ghost}
\end{equation}
\begin{equation}
T_{02}=0,
\end{equation}
\begin{equation}
T_{22}=-\frac{\sqrt{2}\alpha_N}{32\pi}(\overline{\psi}^1\frac{\partial
\psi^1}{\partial
u}-\psi^1\frac{\partial\overline{\psi}^1}{\partial u})=0.
\end{equation}
Let
\begin{equation}
\psi^2=f(v)+ig(v),
\end{equation}
with $f$ and $g$ both real. Thus (\ref{ghost}) can be reduced to
\begin{equation}
fg'=gf',
\end{equation}
where the prime denotes differentiating with respect to $v$. After
a straightforward calculation, there exist three classes of
solutions to the above equation: firstly, $f=0$ and $g$ is an
arbitrary function of $v$; secondly, $f$ is an arbitrary function
of $v$ with $g=0$; finally $g=cf$ where $c$ is an arbitrary
non-zero real constant. Correspondingly, we have three classes of
solutions for $\psi^2$ as follow
\begin{equation}
\psi^2=h(v),
\end{equation}
\begin{equation}
\psi^2=ih(v),
\end{equation}
and
\begin{equation}
\psi^2=(1+ia)h(v),
\end{equation}
where $h$ is an arbitrary real function of $v$, with $a$ an
arbitrary non-zero real constant. Similarly there are three
classes of solutions for $\psi^1$ as follow
\begin{equation}
\psi^1=z(u),
\end{equation}
\begin{equation}
\psi^1=iz(u),
\end{equation}
and
\begin{equation}
\psi^1=(1+ib)z(u),
\end{equation}
where $z$ is an arbitrary real function of $u$, with $b$ an
arbitrary non-zero real constant.

It is obvious that there exist many non-trivial ghost neutrino
solutions in two dimensional flat spacetimes.
\section{Discussion}
We have obtained exhaustive ghost neutrino solutions to
Einstein-Weyl equations for two dimensional flat spacetimes. Due
to conformal invariance of the ghost solutions for two dimensional
case, the exhaustive ghost solutions to Einstein-Weyl equations
for two dimensional curved spacetimes can yield under conformal
transformation. Especially those non-trivial ghost neutrino
solutions present a counterexample to nonsense of two dimensional
analogue for the Einstein-Hilbert action.

We conclude with two interesting problems worthy of further
investigation. Firstly, although it is obvious that there is no
non-trivial solution to gravity-scalar and gravity-electromegnetic
system, there seem to exist non-trivial solutions to gravity-spin
3/2 field coupling system. In addition, up to now, all the ghost
solutions obtained are within the classical framework; it is
interesting to search of ghost states in semi-classical framework,
where matter fields are quantized with gravity treated
classically.
\section*{Acknowledgement}
It is our pleasure to acknowledge Prof. C. Liang for his valuable
discussions and comments. M. Han was partially supported by URPF
from BNU. H. Zhang would like to thank Profs. R. Cai and Y. Zhang
for their hospitality and encouragements at ITP(CAS) where part of
this work was done. In addition, this work was supported in part
by NSFC(grant 10205002).
\section*{Appendix: The Derivation of Einstein-Weyl Equations by Variational Principle}
This appendix serves to present the derivation of Einstein-Weyl
Equations by variational approach, especially the derivation of
stress-energy-momentum tensor for Weyl neutrino field.

Start with the action for Einstein-Weyl system\cite{Wald}
\begin{equation}
S=\int_M\mathcal{L}_{total}\mathbf{e}=\int_M(\mathcal{L}_G+\alpha_N\mathcal{L}_N)\mathbf{e},
\end{equation}
where $\alpha_N$ is an imaginary constant, $\mathcal{L}_G$ is
Lagrangian density for gravitation field, defined as
\begin{equation}
\mathcal{L}_G=R\sqrt{-g},
\end{equation}
and
\begin{equation}
\mathcal{L}_N=\overline{\psi}^{A'}\bigtriangledown_{A'A}\psi^A\sqrt{-g}
\end{equation}
is Lagrangian density for Weyl neutrino field.

The Weyl neutrino equation
\begin{equation}
\bigtriangledown_{A'A}\psi^A=0
\end{equation}
can be easily obtained by variation of $S$ with respect to
$\psi^A$ or $\overline{\psi}^{A'}$. Similarly, variation of $S$
with respect to $g^{ab}$ yields the Einstein equation\cite{Wald}
\begin{equation}
G_{ab}=R_{ab}-\frac{1}{2}Rg_{ab}=8\pi
T_{ab}=8\pi(-\frac{\alpha_N}{8\pi}\frac{1}{\sqrt{-g}}\frac{\delta
S_N}{\delta g^{ab}}),
\end{equation}
where $S_N$ is the action for Weyl neutrino field, given by
\begin{equation}
S_N=\int_M\mathcal{L}_N\mathbf{e}.
\end{equation}

To derive the explicit formula of stress-energy-momentum tensor
for Weyl neutrino
field, we need make some preparations.\\

\textbf{Lemma 1} \emph{For the hybrid vector/spinorial tensor
${\sigma^a}_{AA'}$, there exist the following identities :
\begin{equation}
\sigma_{(aAA'}{\sigma_{b)}}^{AB'}=\frac{1}{2}g_{ab}{\overline{\epsilon}_{A'}}^{B'},\label{lemma1a}
\end{equation}
\begin{equation}
\sigma_{(aAA'}\sigma^{cA'B}\sigma_{b)BB'}={g_{(a}}^c\sigma_{b)AB'}-\frac{1}{2}g_{ab}{\sigma^c}_{AB'},\label{lemma1b}
\end{equation}
where the round brackets denote the symmetrization of $a$ and
$b$.}\\

\textbf{ Proof of (\ref{lemma1a}).}\\
\begin{eqnarray}
\sigma_{(aAA'}{{\sigma_{b)}}^A}_{B'}&=&\frac{1}{2}(\sigma_{aAA'}{{\sigma_b}^A}_{B'}+\sigma_{bAA'}{{\sigma_a}^A}_{B'})\nonumber\\
&=&\frac{1}{2}(\sigma_{aAA'}{{\sigma_b}^A}_{B'}-{{\sigma_b}^A}_{A'}\sigma_{aAB'})\nonumber\\
&=&\frac{1}{2}(\sigma_{aAA'}{{\sigma_b}^A}_{B'}-\sigma_{aAB'}{{\sigma_b}^A}_{A'}),
\end{eqnarray}
which implies $\sigma_{(aAA'}{{\sigma_{b)}}^A}_{B'}$ is
antisymmetric with respect to $A'$ and $B'$. Therefore we have
\begin{equation}
\sigma_{(aAA'}{{\sigma_{b)}}^A}_{B'}=\frac{1}{2}\sigma_{(aAC'}{\sigma_{b)}}^{AC'}\overline{\epsilon}_{A'B'}=\frac{1}{2}g_{ab}\overline{\epsilon}_{A'B'}.
\end{equation}

\textbf{ Proof of (\ref{lemma1b}).}\\
\begin{eqnarray}
\sigma_{(aAA'}\sigma^{cA'B}\sigma_{b)BB'}&=&\frac{1}{2}(\sigma_{aAA'}\sigma^{cA'B}\sigma_{bBB'}+\sigma_{bAA'}\sigma^{cA'B}\sigma_{aBB'})\nonumber\\
&=&\frac{1}{2}(\sigma_{aAA'}\sigma^{cA'B}\sigma_{bBB'}+\sigma_{aBB'}\sigma^{cA'B}\sigma_{bAA'})\nonumber\\
&=&\sigma^{cA'B}\lambda_{ab(AB)(A'B')}+\frac{1}{4}\sigma_{aCC'}\sigma^{cA'B}{\sigma_b}^{CC'}\epsilon_{AB}\overline{\epsilon}_{A'B'}\nonumber\\
&=&\sigma^{cA'B}\lambda_{ab(AB)(A'B')}-\frac{1}{4}g_{ab}{\sigma^c}_{AB'},\label{strange1}
\end{eqnarray}
where $\sigma^{cA'B}\lambda_{ab(AB)(A'B')}$ is given by\cite{Wald}
\begin{eqnarray}
\sigma^{cA'B}\lambda_{ab(AB)(A'B')}&=&\frac{1}{4}(\sigma_{aAA'}\sigma^{cA'B}\sigma_{bBB'}+\sigma_{aBA'}\sigma^{cA'B}\sigma_{bAB'}+\sigma_{aAB'}\sigma^{cA'B}\sigma_{bBA'}+\sigma_{aBB'}\sigma^{cA'B}\sigma_{bAA'})\nonumber\\
&=&\frac{1}{4}(\sigma_{aAA'}\sigma^{cA'B}\sigma_{bBB'}+{g_a}^c\sigma_{bAB'}+{g_b}^c\sigma_{aAB'}+\sigma_{bAA'}\sigma^{cA'B}\sigma_{aBB'})\nonumber\\
&=&\frac{1}{2}\sigma_{(aAA'}\sigma^{cA'B}\sigma_{b)BB'}+\frac{1}{2}{g_{(a}}^c\sigma_{b)AB'}.\label{strange2}
\end{eqnarray}
Thus (\ref{lemma1b}) can yield easily by combining
(\ref{strange1}) and (\ref{strange2}).\\

\textbf{Lemma 2} \emph{Let
\begin{equation}
\widetilde{\bigtriangledown}_a\psi^B=\bigtriangledown_a\psi^B+{{\Theta_a}^B}_C\psi^C,\label{definition}
\end{equation}
where $\widetilde{\bigtriangledown}_a$ and $\bigtriangledown_a$
are covariant derivative operators associated with the metrics
$\widetilde{g}_{ab}$ and $g_{ab}$ respectively; then
${{\Theta_a}^B}_C$ is given by
\begin{equation}
{{\Theta_a}^B}_C=\frac{1}{2}{\widetilde{\sigma}_b}^{BC'}(\bigtriangledown_a{\widetilde{\sigma}^b}_{CC'}+{C^b}_{aCC'}),\label{lemma3}
\end{equation}
where ${\widetilde{\sigma}^a}_{AA'}$ is compatible with
$\widetilde{\bigtriangledown}_a$.}\\

\textbf{Proof.} From (\ref{definition}), it is obvious to yield
\begin{equation}
\widetilde{\bigtriangledown}_a\psi_B=\bigtriangledown_a\psi_B+\Theta_{aBC}\psi^C,
\end{equation}
\begin{equation}
\widetilde{\bigtriangledown}_a\psi^{B'}=\bigtriangledown_a\psi^{B'}+{{\overline{\Theta}_a}^{B'}}_{C'}\psi^{C'},
\end{equation}
where we have employed
\begin{equation}
\widetilde{\bigtriangledown}_a\epsilon_{AB}=\bigtriangledown_a\epsilon_{AB}=0.
\end{equation}
On the other hand
\begin{equation}
\widetilde{\bigtriangledown}_a\epsilon_{AB}=\bigtriangledown_a\epsilon_{AB}+\Theta_{aAC}{\epsilon^C}_B+\Theta_{aBC}{\epsilon_A}^C,
\end{equation}
hereby
\begin{equation}
\Theta_{aAB}=\Theta_{aBA}.\label{symmetry}
\end{equation}
In addition, we have
\begin{eqnarray}
\widetilde{\bigtriangledown}_av^{BB'}&=&\widetilde{\bigtriangledown}_a({\widetilde{\sigma}_b}^{BB'}v^b)={\widetilde{\sigma}_b}^{BB'}\widetilde{\bigtriangledown}_av^b={\widetilde{\sigma}_b}^{BB'}(\bigtriangledown_av^b+{C^b}_{ac}v^c)\nonumber\\
&=&{\widetilde{\sigma}_b}^{BB'}[\bigtriangledown_a({\widetilde{\sigma}^b}_{CC'}v^{CC'})+{C^b}_{ac}v^c]\nonumber\\
&=&\bigtriangledown_av^{BB'}+{\widetilde{\sigma}_b}^{BB'}(\bigtriangledown_a{\widetilde{\sigma}^b}_{CC'}+{C^b}_{aCC'})v^{CC'},\label{tensor}
\end{eqnarray}
on the other hand
\begin{eqnarray}
\widetilde{\bigtriangledown}_av^{BB'}&=&\bigtriangledown_av^{BB'}+{{\Theta_a}^B}_Cv^{CB'}+{{\overline{\Theta}_a}^{B'}}_{C'}v^{BC'}\nonumber\\
&=&\bigtriangledown_av^{BB'}+({{\Theta_a}^B}_C{\overline{\epsilon}_{C'}}^{B'}+{{\overline{\Theta}_a}^{B'}}_{C'}{\epsilon_C}^B)v^{CC'}.\label{spinor}
\end{eqnarray}
Combining (\ref{tensor}) and (\ref{spinor}), we obtain
\begin{equation}
{\widetilde{\sigma}_b}^{BB'}(\bigtriangledown_a{\widetilde{\sigma}^b}_{CC'}+{C^b}_{aCC'})={{\Theta_a}^B}_C{\overline{\epsilon}_{C'}}^{B'}+{{\overline{\Theta}_a}^{B'}}_{C'}{\epsilon_C}^B,
\end{equation}
contracting $B'$ with $C'$ and using (\ref{symmetry}),
(\ref{lemma3}) can yield.\\

According to Lemma 2, with respect to the variation of the metric,
the corresponding variation of ${{\Theta_a}^B}_C$ is given by
\begin{equation}
\delta{{\Theta_a}^B}_C=\frac{1}{2}{\sigma_b}^{BC'}(\bigtriangledown_a\delta{\sigma^b}_{CC'}+\delta{C^b}_{aCC'}),\label{variation1}
\end{equation}
where, for convenience but without loss of generalization, the
variation of ${\sigma^b}_{CC'}$ can be gauge fixed as
\begin{equation}
\delta{\sigma^b}_{CC'}=\frac{1}{2}{\sigma_e}_{CC'}\delta
g^{be},\label{variation2}
\end{equation}
and
\begin{equation}
\delta{C^b}_{aCC'}=\frac{1}{2}{\sigma^c}_{CC'}g^{bd}(\bigtriangledown_a\delta
g_{cd}+\bigtriangledown_c\delta g_{ad}-\bigtriangledown_d\delta
g_{ac}).
\end{equation}
Using
\begin{equation}
g_{ac}\delta g^{bc}=-g^{bc}\delta g_{ac},
\end{equation}
we have
\begin{equation}
\delta{C^b}_{aCC'}=\frac{1}{2}{\sigma_e}_{CC'}(g_{ad}\bigtriangledown^b\delta
g^{ed}-\bigtriangledown_a\delta
g^{be}-g_{ad}\bigtriangledown^e\delta g^{bd}).\label{variation3}
\end{equation}
Unifying (\ref{variation1}), (\ref{variation2}) and
(\ref{variation3}), Corollary 3 can yield. \\

\textbf{Corollary 3} \emph{The final version for the variation of
${{\Theta_a}^B}_C$ is given by
\begin{equation}
\delta{{\Theta_a}^B}_C=\frac{1}{4}{\sigma_b}^{BC'}{\sigma_e}_{CC'}g_{ad}(\bigtriangledown^b\delta
g^{ed}-\bigtriangledown^e\delta g^{bd}).
\end{equation}}\\

With the above preparations, we can derive the
stress-energy-momentum tensor for Weyl neutrino field.\\

\textbf{Theorem 4} \emph{The stress-energy-momentum tensor for
Weyl neutrino field is given by
\begin{equation}
T_{ab}=-\frac{\alpha_N}{32\pi}(\overline{\psi}^{A'}\sigma_{(aA'A}\bigtriangledown_{b)}\psi^A-\bigtriangledown_{(a}\overline{\psi}^{A'}\sigma_{b)A'A}\psi^A),
\end{equation}
where the round brackets denote the symmetrization of $a$ and
$b$.}\\

\textbf{Proof.} Start with
\begin{equation}
\delta S_N=\int_M
\delta\mathcal{L}_N\mathbf{e}=\int_M(\overline{\psi}^{A'}\delta{\sigma^a}_{A'A}\bigtriangledown_a\psi^A\sqrt{-g}+\overline{\psi}^{A'}{\sigma^a}_{A'A}\delta{{\Theta_a}^A}_B\psi^B\sqrt{-g}+\overline{\psi}^{A'}\bigtriangledown_{A'A}\psi^A\delta\sqrt{-g})\mathbf{e},
\end{equation}
using (\ref{variation2}), the first integrand gives
\begin{equation}
\frac{1}{2}\overline{\psi}^{A'}\sigma_{(aA'A}\bigtriangledown_{b)}\psi^A\delta
g^{ab}\sqrt{-g},\label{action1}
\end{equation}
similarly, employing Corollary 3 and making a partial integration,
the second term in the integrand yields
\begin{equation}
\frac{1}{4}[
\sigma_{(aBB'}{\sigma_{b)}}^{BC'}\bigtriangledown_{CC'}(\overline{\psi}^{B'}\psi^C)-
\sigma_{(aCC'}\sigma^{cC'B}\sigma_{b)BB'}\bigtriangledown_c(\overline{\psi}^{B'}\psi^C)
]\delta g^{ab}\sqrt{-g},
\end{equation}
using Lemma 1 and the Weyl neutrino field equation, which can be
written as
\begin{equation}
-\frac{1}{4}(\overline{\psi}^{A'}\sigma_{(aA'A}\bigtriangledown_{b)}\psi^A+\bigtriangledown_{(a}\overline{\psi}^{A'}\sigma_{b)A'A}\psi^A)\delta
g^{ab}\sqrt{-g},\label{action2}
\end{equation}
the third integrand vanishes by virtue of the Weyl neutrino field
equation.

Thus, combining (\ref{action1}) and (\ref{action2}), we have
\begin{equation}
\frac{\delta S_N}{\delta
g^{ab}}=\frac{1}{4}(\overline{\psi}^{A'}\sigma_{(aA'A}\bigtriangledown_{b)}\psi^A-\bigtriangledown_{(a}\overline{\psi}^{A'}\sigma_{b)A'A}\psi^A)\sqrt{-g},
\end{equation}
therefore
\begin{equation}
T_{ab}=-\frac{\alpha_N}{32\pi}(\overline{\psi}^{A'}\sigma_{(aA'A}\bigtriangledown_{b)}\psi^A-\bigtriangledown_{(a}\overline{\psi}^{A'}\sigma_{b)A'A}\psi^A).
\end{equation}

\end{document}